\pdfoutput=1
\documentclass[11pt]{article}

\usepackage[margin=1in]{geometry}
\usepackage{amsmath,amssymb}
\usepackage{graphicx}
\usepackage{booktabs}
\usepackage{caption}
\usepackage[numbers,sort&compress]{natbib}
\usepackage[colorlinks=true,linkcolor=blue,citecolor=blue,urlcolor=blue]{hyperref}
\usepackage{placeins}
\usepackage{xcolor}
\usepackage{array}
\newcommand{\bhNVDA}{126.1\%}
\newcommand{\bhJPM}{-3.7\%}

\newcommand{\lgbmSeedCv}{0.9\%}

\title{\textbf{Does News Sentiment Add to Return Forecasts? A Leakage-Controlled Walk-Forward Evaluation with Market-Aligned Labels}}
\author{Sohom Mandal\thanks{Corresponding author: \href{mailto:s.mandal@northeastern.edu}{s.mandal@northeastern.edu}}\\[2pt]
\small Professor of Analytics, Northeastern University, Toronto\\
\small MBA Candidate, Rotman School of Management, University of Toronto}
\date{November 30, 2026}

\begin{document}
\maketitle

\begin{abstract}
Does news sentiment help predict stock returns? This paper tests a simple idea on market data: add a causal headline-sentiment signal to a standard set of price-based features and check, under strict out-of-sample rules, whether next-day return forecasts improve. Sentiment is zero when there is no news and otherwise carried forward with exponential decay after a one-day publication lag, so it never uses future information. Using 4,843 headlines and daily prices for NVIDIA (NVDA) and JPMorgan Chase (JPM), every threshold, scaler and hyperparameter is fitted on training data only, and models are evaluated on expanding walk-forward windows against a random-walk forecast, a historical mean, AR(1), ridge regression and LightGBM. Over 476 out-of-sample days in 2019--2020, no model beats the zero-return forecast with statistical confidence on the pooled sample, and adding sentiment changes mean squared error by less than 0.6\% in either direction (all $p>0.6$). The only significant gain over the random walk is for NVDA with linear models, concentrated in the 2020 market crash. A long/short rule on the ridge forecasts earns a large return for NVDA over a short, unusual period but loses money for JPM. On a three-class market-impact label, a majority-class rule reaches 74\% accuracy but only 0.33 balanced accuracy, showing why accuracy alone is misleading. The pipeline is short, fully reproducible, and ends with concrete extensions.
\end{abstract}

\section{Introduction}
Stock return prediction is hard, and published improvements from news sentiment are often small. Small gains are easily produced by design choices that are invisible in a results table: smoothing a sentiment series with a method that uses future observations, forecasting a non-stationary price level (where yesterday's price is already an excellent forecast), choosing label thresholds with the whole sample, or comparing against weak baselines \citep{welch2008}. The aim of this paper is to measure, as cleanly as the available data allow, how much a headline-based sentiment signal adds to a return forecast once these sources of optimism are removed.

The paper makes four contributions.
\begin{enumerate}
\item \textbf{A causal sentiment construction.} Sentiment is zero without news and otherwise carried forward with exponential decay after a publication lag, so no value at time $t$ depends on any later observation. Interpolating across non-news days would not be causal, because it generally draws on both earlier and later sentiment and so lets future news influence past values.
\item \textbf{A complete, training-fitted market-impact label.} Labels are based on the abnormal three-day return (stock minus sector benchmark) with a three-class rule that covers every observation and a threshold estimated on training data only.
\item \textbf{A strict walk-forward protocol with significance tests.} Forecasts are compared with a random-walk benchmark, a historical-mean forecast, AR(1), ridge regression and LightGBM using Diebold--Mariano tests and paired block-bootstrap confidence intervals, with results by fold and by random seed.
\item \textbf{Evidence on what the signal is worth.} For these two stocks and this period, the sentiment signal adds nothing detectable to next-day return forecasts, and the classification task is dominated by one class.
\end{enumerate}
\section{Related Work}
\textbf{Return forecasting.} Classical time-series models such as ARIMA handle linear structure \citep{mehtab2019}, and recurrent networks capture longer dependencies \citep{rather2015}. Out-of-sample return predictability is notoriously weak and often fails to beat a simple historical mean \citep{welch2008}, which motivates our simple baselines.

\textbf{Financial sentiment.} Lexicon methods such as VADER \citep{hutto2014} and the Loughran--McDonald word lists \citep{loughran2011} are fast but context-blind. Fine-tuned transformers such as FinBERT variants \citep{gssi2023} are more accurate on benchmark sentiment tasks.

\textbf{Market-aligned labels.} Instead of human annotation, a news item can be labelled by the market's subsequent response. The FNSPID corpus \citep{dong2024} offers time-stamped news paired with prices. Because such a label is built from future returns, it must be fitted on training data only, and what a model learns is closer to market-impact classification than to linguistic tone.

\section{Data and Timing}
\label{sec:data}
\textbf{Prices.} Daily adjusted OHLCV for NVDA and JPM, together with SPY (market), SOXX (semiconductor benchmark for NVDA) and XLF (financials benchmark for JPM), cover 1 December 2016 to 29 December 2023 (1,781 sessions each).

\textbf{News.} Headlines come from the FNSPID corpus \citep{dong2024}: 4,843 items after filtering and de-duplication (NVDA 2,021; JPM 2,822), of which 36.5\% explicitly mention the company. Duplicates are removed by URL and then by ticker, normalized title and session date. The audit files in the repository document each filtering step.

\textbf{Coverage limits.} News coverage ends in June 2020 (last item dated 11 June 2020), JPM news begins in September 2018, and 24.0\% (JPM) and 36.8\% (NVDA) of trading days have at least one item. Experiments are therefore restricted to the window in which news exists, 17 September 2018 to 10 June 2020 for JPM and 1 December 2016 to 10 June 2020 for NVDA. The 2021--2023 price history is unused: evaluating a sentiment model there would mean feeding it empty inputs.

\textbf{Timing rules.} Only 0.4\% of items carry a time of day, and 0.1\% are published after the 16:00 close. Items published at or after 16:00 ET, on weekends or on holidays are assigned to the next trading session, and date-only items to their own date's session if it is a trading day. Because intraday timing is unknown for almost all items, a news item dated session $s$ is treated as unavailable until the close of session $s+1$ (one-day lag). This conservative rule removes the possibility that a same-day, after-close item informs a forecast made at that day's close. Multiple items on one session are averaged. Mixed timing makes this a limitation rather than a solved problem, and Section~\ref{sec:limits} discusses it.

\section{Methodology}
\subsection{Targets and Features}
For stock $i$ and session $t$ the forecast target is the next-day return
\begin{equation}
r_{t+1}=\frac{P_{t+1}-P_t}{P_t}.
\end{equation}
Predicted prices for plotting are reconstructed as $\hat P_{t+1}=P_t(1+\hat r_{t+1})$. Features at $t$ use only data up to $t$: returns over 1, 2, 3 and 5 days, 5- and 20-day return volatility, intraday range, open-to-close return, log volume change and the market (SPY) return. Technical-only models use these ten features. Sentiment models add two more, the causal sentiment score and the log of one plus the (lagged) article count.

\subsection{Causal Sentiment}
Each headline receives a VADER compound score in $[-1,1]$ \citep{hutto2014}, averaged within session. With publication lag $\ell=1$ and decay factor $\lambda=0.5^{1/h}$ (half-life $h=3$ sessions),
\begin{equation}
s_t=\begin{cases}\bar a_{t-\ell} & \text{if news were dated session } t-\ell,\\ \lambda\, s_{t-1} & \text{otherwise,}\end{cases}\qquad s_0=0.
\end{equation}
Thus $s_t=0$ until the first news item arrives, and $s_t$ never uses information dated after $t-\ell$. A spline would not be causal here, because it fits a curve through known values on both sides of a gap, so a later news score would shape earlier sentiment. Any other sentiment score, such as FinBERT, can be substituted in the same way.

\subsection{Market-Impact Labels}
The abnormal three-day return is
\begin{equation}
AR_{i,t}=\frac{P_{i,t+3}-P_{i,t}}{P_{i,t}}-\frac{B_{t+3}-B_t}{B_t},
\end{equation}
where $B$ is SOXX for NVDA and XLF for JPM. With $\theta=k\,\hat\sigma_{AR}^{\text{train}}$ and $k=1$, where $\hat\sigma^{\text{train}}_{AR}$ is the standard deviation of $AR$ in the training window,
\begin{equation}
\text{Label}_{i,t}=\begin{cases}\text{Positive} & AR_{i,t}>\theta\\ \text{Negative} & AR_{i,t}<-\theta\\ \text{Neutral} & \text{otherwise.}\end{cases}
\end{equation}
The rule covers every observation, so none is discarded.

\subsection{Models}
All models forecast $r_{t+1}$ and are refit in every fold on training rows only. The benchmark set is: (i) the \textbf{random walk}, forecast zero return; (ii) the \textbf{historical mean} of training returns; (iii) \textbf{AR(1)} on daily returns, with parameters fixed from training data; (iv) \textbf{ridge regression} \citep{hoerl1970} with standardized features (scaler fitted on training rows) and the penalty chosen by time-series cross-validation inside the training window; and (v) \textbf{LightGBM} \citep{ke2017} with fixed shallow trees (200 trees, learning rate 0.03, 7 leaves) averaged over five seeds. Ridge and LightGBM are each run with and without the two sentiment features. The final three training rows of each fold are purged because their three-day targets overlap the test window.

\subsection{Evaluation}
Test windows are expanding walk-forward folds: training through 28 June 2019 with testing 1 July--31 December 2019, and training through 31 December 2019 with testing 2 January--10 June 2020. All boundaries are trading days. Every table is computed from one prediction file, and a program check asserts that $\text{RMSE}=\sqrt{\text{MSE}}$ for every row before a table is written. Beyond MSE, RMSE and MAE we report directional accuracy, the correlation of forecasts with realized returns, and precision for positive and negative predicted moves. Differences in squared error against a benchmark are tested with a Diebold--Mariano test \citep{dm1995} using the Harvey--Leybourne--Newbold small-sample correction \citep{hln1997} and with a 95\% paired moving-block bootstrap interval \citep{kunsch1989} (block length 10, 2,000 resamples). Classification is evaluated with accuracy, balanced accuracy, macro- and weighted-F1, per-class scores and a confusion matrix, against a majority-class baseline. The trading test goes long or short by the sign of the forecast, charges 5 basis points of one-way cost on every position change and reports cumulative and annualized return, Sharpe ratio, maximum drawdown, turnover and number of trades.

\begin{figure}[!htb]\centering
\includegraphics[width=\textwidth]{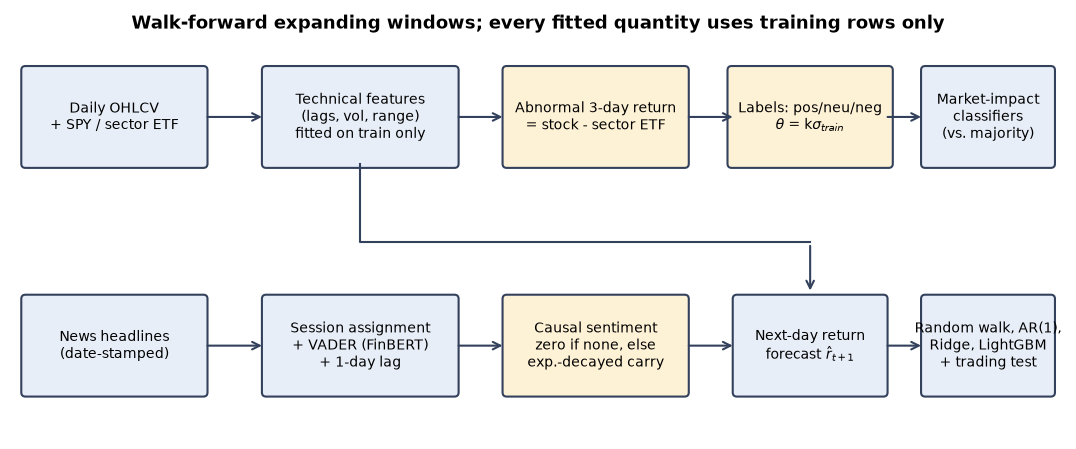}
\caption{Pipeline. Yellow boxes are the new steps (causal sentiment and training-fitted abnormal-return labels); every fitted quantity uses training rows only.}
\label{fig:arch}
\end{figure}

\section{Results}
\subsection{Label Distribution}
Table~\ref{tab:labels} shows the training-fitted thresholds and class counts. Neutral dominates every window, with one test window (NVDA, 2019-H2) containing no negative cases at all. Any classifier is therefore judged against a strong majority-class baseline, and class balance differs between training and test periods.
\begin{table}[!htb]\centering\small\begin{tabular}{llrrrrr}\toprule Stock & Test window & Threshold $\theta$ & Train $n$ & Train neg/neu/pos & Test $n$ & Test neg/neu/pos \\\midrule
NVDA & 2019-H2 & 3.45\% & 624 & 65/504/55 & 128 & 0/117/11 \\
NVDA & 2020-H1 & 3.27\% & 752 & 73/604/75 & 108 & 10/74/24 \\
JPM & 2019-H2 & 0.90\% & 174 & 22/129/23 & 128 & 6/100/22 \\
JPM & 2020-H1 & 0.86\% & 302 & 30/217/55 & 108 & 28/60/20 \\
\bottomrule\end{tabular}\caption{Training-fitted threshold $\theta$ and class counts (negative/neutral/positive) by fold.}\label{tab:labels}\end{table}
\begin{figure}[!htb]\centering\includegraphics[width=.85\textwidth]{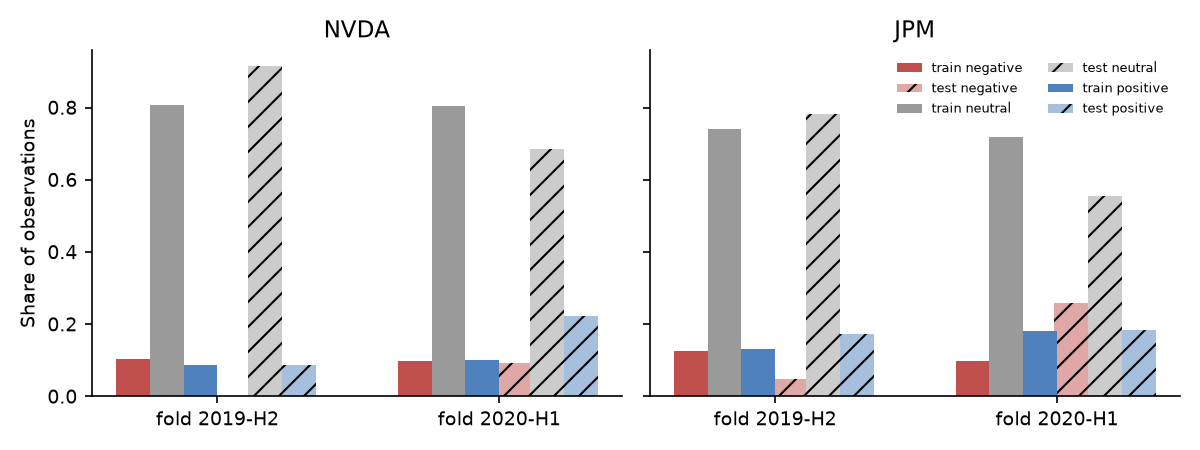}\caption{Label distribution by stock and fold. Solid bars are training, hatched bars are test.}\end{figure}

\subsection{Return Forecasting}
Table~\ref{tab:returns} reports the 476 pooled out-of-sample forecasts. Every model has an RMSE within about 1\% of the random walk. Ridge forecasts are marginally ahead of the random walk and have a positive correlation with realized returns; LightGBM is marginally behind. Directional accuracy lies between 47\% and 54\%, close to a coin flip. Percentage price errors are not used, because predicting returns rather than price levels avoids the artificially small errors that come from consecutive prices being close.
\begin{table}[!htb]\centering\scriptsize\resizebox{\textwidth}{!}{\begin{tabular}{lrrrrrrr}\toprule Model & MSE$\times10^{4}$ & RMSE & MAE & Dir.\ acc. & Corr. & Prec.\ up & Prec.\ down \\

\multicolumn{8}{l}{\textit{NVDA}} \\
Random walk (zero return) & 12.685 & 0.03562 & 0.02458 & -- & -- & -- & -- \\
Historical mean & 12.596 & 0.03549 & 0.02447 & 0.555 & 0.028 & 0.555 & -- \\
AR(1) & 12.412 & 0.03523 & 0.02426 & 0.580 & 0.287 & 0.574 & 0.636 \\
Ridge (technical) & 12.464 & 0.03530 & 0.02434 & 0.571 & 0.356 & 0.564 & 1.000 \\
Ridge (technical + sentiment) & 12.462 & 0.03530 & 0.02434 & 0.571 & 0.352 & 0.564 & 1.000 \\
LightGBM (technical) & 12.728 & 0.03568 & 0.02483 & 0.466 & 0.092 & 0.522 & 0.394 \\
LightGBM (technical + sentiment) & 12.801 & 0.03578 & 0.02477 & 0.496 & 0.075 & 0.548 & 0.427 \\
\midrule
\multicolumn{8}{l}{\textit{JPM}} \\
Random walk (zero return) & 10.131 & 0.03183 & 0.01936 & -- & -- & -- & -- \\
Historical mean & 10.145 & 0.03185 & 0.01938 & 0.525 & -0.052 & 0.525 & -- \\
AR(1) & 10.245 & 0.03201 & 0.01949 & 0.437 & -0.216 & 0.478 & 0.200 \\
Ridge (technical) & 10.191 & 0.03192 & 0.01942 & 0.517 & -0.042 & 0.522 & 0.333 \\
Ridge (technical + sentiment) & 10.182 & 0.03191 & 0.01941 & 0.508 & -0.066 & 0.517 & 0.000 \\
LightGBM (technical) & 10.383 & 0.03222 & 0.01996 & 0.479 & -0.029 & 0.509 & 0.453 \\
LightGBM (technical + sentiment) & 10.353 & 0.03218 & 0.01983 & 0.487 & -0.020 & 0.519 & 0.462 \\
\midrule
\multicolumn{8}{l}{\textit{Both stocks pooled}} \\
Random walk (zero return) & 11.408 & 0.03378 & 0.02197 & -- & -- & -- & -- \\
Historical mean & 11.370 & 0.03372 & 0.02193 & 0.540 & 0.023 & 0.540 & -- \\
AR(1) & 11.329 & 0.03366 & 0.02187 & 0.508 & 0.088 & 0.527 & 0.368 \\
Ridge (technical) & 11.327 & 0.03366 & 0.02188 & 0.544 & 0.075 & 0.543 & 0.600 \\
Ridge (technical + sentiment) & 11.322 & 0.03365 & 0.02188 & 0.540 & 0.104 & 0.541 & 0.500 \\
LightGBM (technical) & 11.555 & 0.03399 & 0.02239 & 0.473 & 0.056 & 0.516 & 0.427 \\
LightGBM (technical + sentiment) & 11.577 & 0.03402 & 0.02230 & 0.492 & 0.047 & 0.535 & 0.446 \\
\bottomrule\end{tabular}}\caption{Next-day return forecasts, out-of-sample (both folds combined). Directional accuracy and precision are undefined for the zero forecast. LightGBM figures use the mean forecast of five seeds.}\label{tab:returns}\end{table}
\begin{figure}[!htb]\centering\includegraphics[width=.9\textwidth]{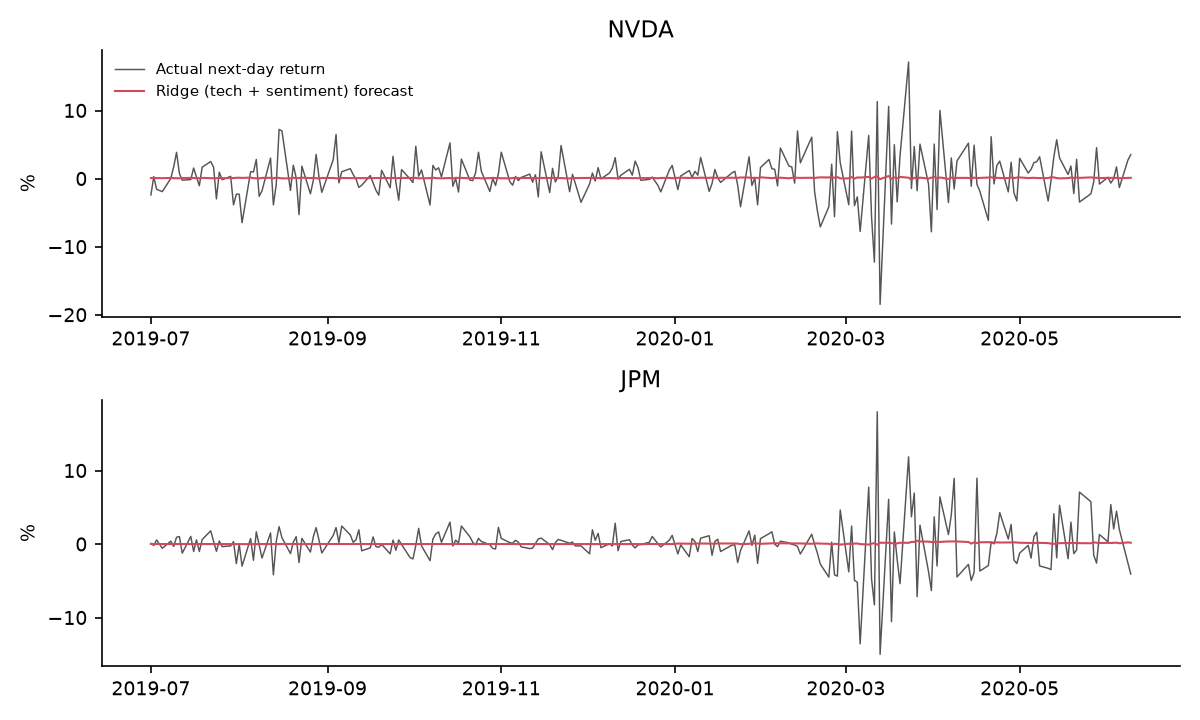}\caption{Actual next-day returns and ridge (technical + sentiment) forecasts. The forecasts are far smaller than the realized moves, as expected when predictability is weak.}\end{figure}

\subsection{Walk-Forward Performance and Seeds}
\begin{figure}[!htb]\centering\includegraphics[width=.9\textwidth]{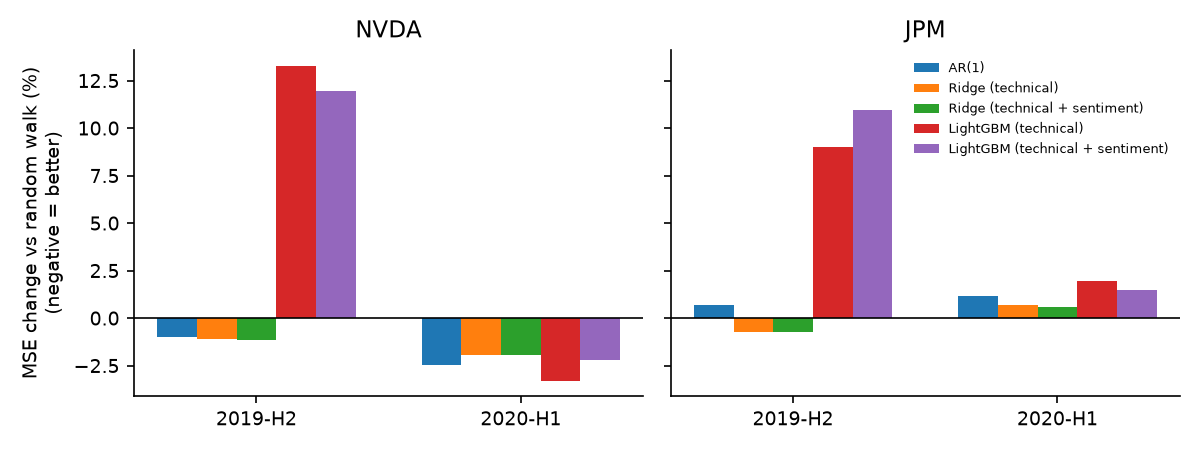}\caption{Change in MSE relative to the random walk by fold.}\label{fig:wf}\end{figure}

Table~\ref{tab:folds} and Figure~\ref{fig:wf} show that the linear models' gains for NVDA appear in both windows but are largest in 2020-H1, which includes the March 2020 crash with extreme negative return autocorrelation; for JPM none of the models beats the random walk in 2020-H1. LightGBM is clearly worse than the random walk in the calm 2019-H2 window. Across five seeds the RMSE of LightGBM has a coefficient of variation of at most \lgbmSeedCv{} in any stock-window cell, so seed noise is small relative to the gaps between models. In all four stock-window cells, adding sentiment changes ridge MSE relative to the random walk by at most 0.1 percentage points.
\begin{table}[!htb]\centering\scriptsize\resizebox{\textwidth}{!}{\begin{tabular}{llrrrrrr}\toprule Stock & Test window & RW MSE$\times10^{4}$ & AR(1) & Ridge tech & Ridge +sent & LGBM tech & LGBM +sent \\\midrule
NVDA & 2019-H2 & 5.155 & -1.00\% & -1.10\% & -1.12\% & +13.30\% & +11.96\% \\
NVDA & 2020-H1 & 21.446 & -2.47\% & -1.93\% & -1.93\% & -3.28\% & -2.17\% \\
JPM & 2019-H2 & 1.418 & +0.73\% & -0.73\% & -0.70\% & +9.01\% & +10.97\% \\
JPM & 2020-H1 & 20.269 & +1.16\% & +0.71\% & +0.61\% & +1.96\% & +1.48\% \\
\bottomrule\end{tabular}}\caption{MSE of each model relative to the random walk (negative is better), by walk-forward window.}\label{tab:folds}\end{table}

\subsection{Statistical Significance}
Table~\ref{tab:sig} gives paired tests. Pooled across both stocks, no model differs significantly from the zero-return forecast. For NVDA alone, AR(1) and ridge (with or without sentiment) are significantly better than the random walk ($p\approx0.01$); this is a single stock over two windows with an unusual crash in the second, and no multiple-comparison correction is applied, so it should be read as suggestive. The key test for the paper's question compares each model with and without sentiment: the changes in MSE are below 0.6\% in absolute value, the confidence intervals straddle zero and every $p$-value exceeds 0.6.
\begin{table}[!htb]\centering\scriptsize\resizebox{\textwidth}{!}{\begin{tabular}{llrrrr}\toprule Comparison & Stock & $\Delta$MSE & 95\% block-bootstrap CI of $\Delta$sq.\ error ($\times10^{4}$) & DM & $p$ \\\midrule
AR(1) vs.\ random walk & NVDA & -2.15\% & [-0.573, -0.077] & -2.61 & 0.010 \\
AR(1) vs.\ random walk & JPM & +1.13\% & [-0.019, +0.321] & +1.79 & 0.075 \\
AR(1) vs.\ random walk & ALL & -0.69\% & [-0.252, +0.082] & -1.28 & 0.202 \\
Ridge (technical) vs.\ random walk & NVDA & -1.74\% & [-0.426, -0.084] & -2.68 & 0.008 \\
Ridge (technical) vs.\ random walk & JPM & +0.60\% & [-0.086, +0.234] & +0.55 & 0.584 \\
Ridge (technical) vs.\ random walk & ALL & -0.70\% & [-0.202, +0.051] & -1.16 & 0.247 \\
Ridge (technical + sentiment) vs.\ random walk & NVDA & -1.75\% & [-0.429, -0.083] & -2.68 & 0.008 \\
Ridge (technical + sentiment) vs.\ random walk & JPM & +0.51\% & [-0.064, +0.192] & +0.64 & 0.522 \\
Ridge (technical + sentiment) vs.\ random walk & ALL & -0.75\% & [-0.203, +0.032] & -1.48 & 0.141 \\
LightGBM (technical + sentiment) vs.\ random walk & NVDA & +0.92\% & [-0.757, +0.860] & +0.25 & 0.800 \\
LightGBM (technical + sentiment) vs.\ random walk & JPM & +2.19\% & [-0.067, +0.403] & +1.27 & 0.207 \\
LightGBM (technical + sentiment) vs.\ random walk & ALL & +1.48\% & [-0.285, +0.554] & +0.69 & 0.490 \\
Ridge (technical + sentiment) vs.\ Ridge (technical) & NVDA & -0.01\% & [-0.007, +0.005] & -0.35 & 0.729 \\
Ridge (technical + sentiment) vs.\ Ridge (technical) & JPM & -0.09\% & [-0.046, +0.027] & -0.29 & 0.773 \\
Ridge (technical + sentiment) vs.\ Ridge (technical) & ALL & -0.05\% & [-0.024, +0.011] & -0.32 & 0.748 \\
LightGBM (technical + sentiment) vs.\ LightGBM (technical) & NVDA & +0.57\% & [-0.114, +0.296] & +0.50 & 0.616 \\
LightGBM (technical + sentiment) vs.\ LightGBM (technical) & JPM & -0.29\% & [-0.137, +0.075] & -0.51 & 0.609 \\
LightGBM (technical + sentiment) vs.\ LightGBM (technical) & ALL & +0.18\% & [-0.085, +0.145] & +0.27 & 0.784 \\
\bottomrule\end{tabular}}\caption{Diebold--Mariano tests and block-bootstrap intervals for the difference in squared forecast error (alternative minus benchmark; negative favours the model listed first).}\label{tab:sig}\end{table}
\begin{figure}[!htb]\centering\includegraphics[width=.85\textwidth]{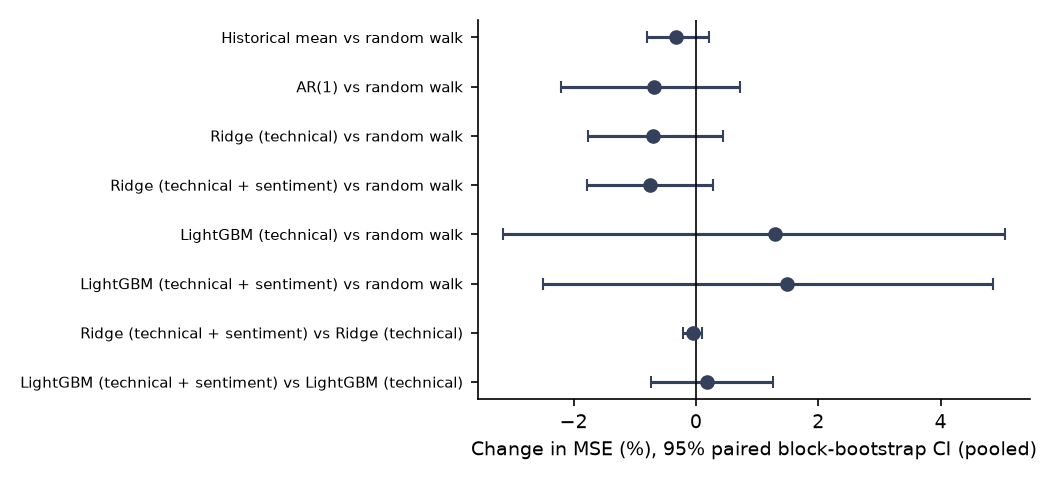}\caption{Percentage change in pooled MSE with 95\% block-bootstrap intervals.}\end{figure}

\subsection{Market-Impact Classification}
Table~\ref{tab:class} compares simple classifiers on the three-class abnormal-return label. The majority-class rule reaches the highest accuracy and weighted F1 on the pooled sample, but its balanced accuracy is exactly one third and its macro-F1 is the lowest of the machine-learning baselines. Logistic regression with sentiment is slightly better than without on balanced accuracy and macro-F1, but all values remain close to chance level. A direct sign rule on the VADER score is worse than random. Any more elaborate classifier should therefore be judged on balanced accuracy and macro-F1 against these reference values, not on accuracy.
\begin{table}[!htb]\centering\scriptsize\resizebox{\textwidth}{!}{\begin{tabular}{llrrrrr}\toprule Stock & Model & Acc. & Bal.\ acc. & Macro-F1 & Weighted-F1 & F1 neg/neu/pos \\\midrule
NVDA & Majority class & 0.809 & 0.333 & 0.298 & 0.724 & 0.00/0.89/0.00 \\
NVDA & VADER sign rule & 0.326 & 0.314 & 0.241 & 0.397 & 0.07/0.45/0.20 \\
NVDA & Logistic (technical) & 0.479 & 0.354 & 0.314 & 0.555 & 0.06/0.64/0.24 \\
NVDA & Logistic (technical + sentiment) & 0.475 & 0.430 & 0.344 & 0.553 & 0.09/0.62/0.32 \\
\midrule
JPM & Majority class & 0.678 & 0.333 & 0.269 & 0.548 & 0.00/0.81/0.00 \\
JPM & VADER sign rule & 0.254 & 0.300 & 0.241 & 0.271 & 0.20/0.30/0.22 \\
JPM & Logistic (technical) & 0.326 & 0.343 & 0.290 & 0.357 & 0.14/0.41/0.32 \\
JPM & Logistic (technical + sentiment) & 0.347 & 0.330 & 0.299 & 0.384 & 0.13/0.46/0.31 \\
\midrule
Pooled & Majority class & 0.744 & 0.333 & 0.284 & 0.634 & 0.00/0.85/0.00 \\
Pooled & VADER sign rule & 0.290 & 0.308 & 0.245 & 0.332 & 0.14/0.38/0.21 \\
Pooled & Logistic (technical) & 0.403 & 0.355 & 0.313 & 0.460 & 0.11/0.54/0.29 \\
Pooled & Logistic (technical + sentiment) & 0.411 & 0.372 & 0.325 & 0.468 & 0.12/0.55/0.31 \\
\bottomrule\end{tabular}}\caption{Classification of three-day abnormal-return labels (test windows only).}\label{tab:class}\end{table}
\begin{figure}[!htb]\centering\includegraphics[width=.8\textwidth]{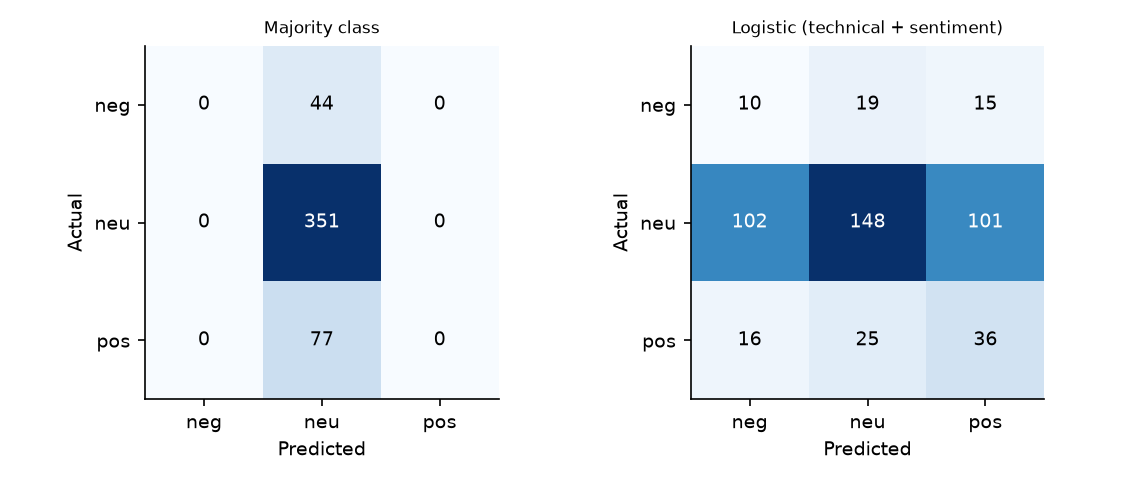}\caption{Pooled confusion matrices for the majority-class rule and logistic regression with sentiment.}\end{figure}

\subsection{Trading Test}
Table~\ref{tab:trading} reports a simple long/short rule on forecast sign after costs. Buy-and-hold over the same test days returned \bhNVDA{} for NVDA and \bhJPM{} for JPM. The ridge and AR(1) rules perform very well for NVDA, but the entire gain arises in March--June 2020, when markets were unusually mean-reverting, and the same rules lose money for JPM. LightGBM, which trades much more often, loses money for both stocks. Given the short sample, one dominant episode, and no significance on the pooled sample, we do not interpret these figures as evidence of a tradable strategy.
\begin{table}[!htb]\centering\scriptsize\resizebox{\textwidth}{!}{\begin{tabular}{llrrrrrrr}\toprule Stock & Model & Cum.\ ret. & Ann.\ ret. & Sharpe & Max DD & Turnover & Trades & Costs paid \\\midrule
NVDA & Historical mean & 126.0\% & 137.1\% & 1.82 & -37.6\% & 0.004 & 1 & 0.05\% \\
NVDA & AR(1) & 318.1\% & 354.8\% & 3.01 & -19.6\% & 0.357 & 43 & 4.25\% \\
NVDA & Ridge (technical) & 310.7\% & 346.3\% & 2.98 & -31.2\% & 0.071 & 9 & 0.85\% \\
NVDA & Ridge (technical + sentiment) & 310.7\% & 346.3\% & 2.98 & -31.2\% & 0.071 & 9 & 0.85\% \\
NVDA & LightGBM (technical) & -24.2\% & -25.5\% & -0.24 & -41.7\% & 0.937 & 112 & 11.15\% \\
NVDA & LightGBM (technical + sentiment) & -17.7\% & -18.6\% & -0.08 & -37.5\% & 0.887 & 106 & 10.55\% \\
\midrule
JPM & Historical mean & -3.7\% & -4.0\% & 0.17 & -43.6\% & 0.004 & 1 & 0.05\% \\
JPM & AR(1) & -67.6\% & -69.7\% & -2.11 & -75.0\% & 0.525 & 63 & 6.25\% \\
JPM & Ridge (technical) & -31.5\% & -33.0\% & -0.53 & -59.9\% & 0.088 & 11 & 1.05\% \\
JPM & Ridge (technical + sentiment) & -48.6\% & -50.6\% & -1.13 & -69.9\% & 0.071 & 9 & 0.85\% \\
JPM & LightGBM (technical) & -44.8\% & -46.7\% & -0.99 & -45.1\% & 0.937 & 112 & 11.15\% \\
JPM & LightGBM (technical + sentiment) & -22.4\% & -23.6\% & -0.28 & -42.9\% & 0.803 & 96 & 9.55\% \\
\bottomrule\end{tabular}}\caption{Long/short rule on forecast sign, 5 bp one-way cost.}\label{tab:trading}\end{table}
\begin{figure}[!htb]\centering\includegraphics[width=.95\textwidth]{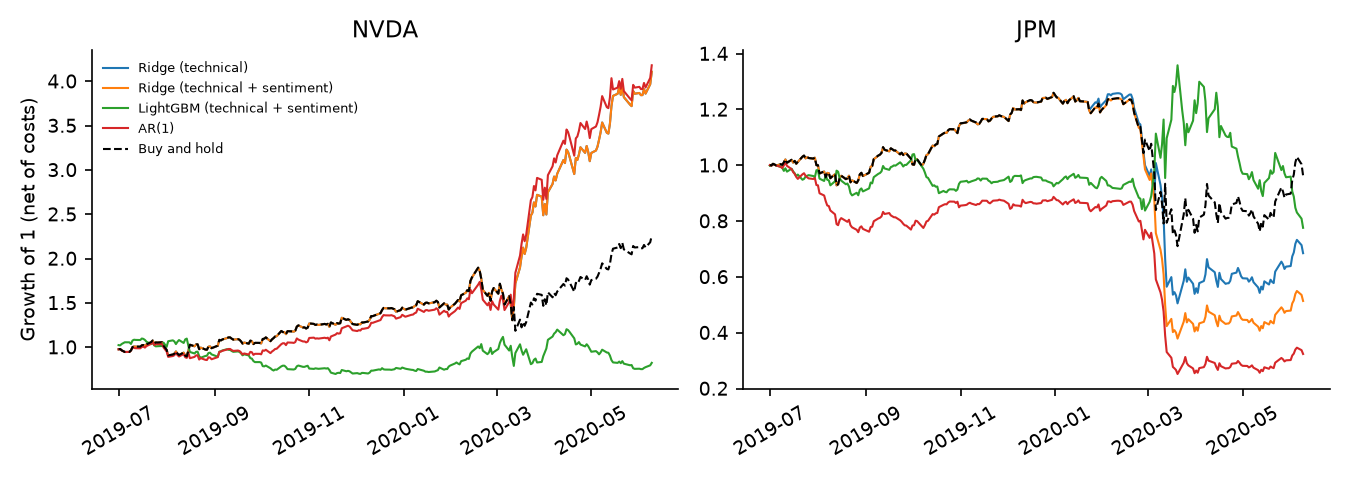}\caption{Cumulative return of the trading rules net of costs.}\end{figure}

\FloatBarrier
\section{Discussion}
The headline finding is negative: in this sample, VADER-based causal headline sentiment does not measurably improve next-day return forecasts. Several explanations are consistent with the evidence. Headline sentiment from a lexicon is a weak signal. Only about a quarter to a third of days carry any news, so the series is mostly decayed carry-forward. The one-day lag, needed because intraday timing is unavailable, discards the fastest reaction. And next-day returns of large, liquid stocks are close to unpredictable. The results do not show that richer sentiment, such as a transformer score or full article text, would also fail.

The contrast between accuracy and balanced accuracy in the classification task is a useful warning for work on market-aligned labels. When one class makes up three quarters of the observations, high accuracy and moderate weighted F1 are what a classifier that always answers ``neutral'' achieves. Reporting macro-F1, balanced accuracy and per-class scores is necessary for any claim of improvement.

\section{Limitations}
\label{sec:limits}
\begin{itemize}
\item \textbf{Scope.} Two stocks and two test windows (238 days per stock). Generalization to other sectors needs more tickers with news coverage beyond 2020.
\item \textbf{Regime.} Both windows are dominated by the 2020 pandemic shock in the second half. Results for 2021--2023 cannot be produced without a news source for those years.
\item \textbf{Timing.} 99.6\% of headlines have only a date. The one-day lag is conservative, but a more precise treatment of after-close news is impossible with this data.
\item \textbf{Sentiment measure.} Only VADER on headlines is used; FinBERT and full article text may behave differently.
\item \textbf{Multiple tests.} The NVDA-only significance is not corrected for multiple comparisons.
\item \textbf{Fixed choices.} Label threshold ($k=1$), horizon (3 days), decay half-life (3 sessions) and the one-day lag were fixed in advance, not tuned.
\item \textbf{Trading.} The strategy is a stylized sign rule without borrow costs, slippage or position sizing.
\end{itemize}

\section{Conclusion}
Under a leakage-controlled protocol, adding causal VADER headline sentiment to technical features does not improve next-day return forecasts for NVDA or JPM beyond sampling noise, and no model beats a random walk with statistical confidence on the pooled data. The main practical results are methodological: use causal sentiment, fit every parameter on training data, forecast returns instead of prices, compare against simple baselines, report class-level metrics next to accuracy, and test significance. 

\section{Future Work}
\begin{itemize}
\item Add more stocks across sectors and a news source that extends beyond mid-2020, so that walk-forward folds can cover 2021--2023 and different market regimes.
\item Replace VADER with a finance-specific language-model score (for example FinBERT) and compare it with the lexicon score.
\item Run a sensitivity analysis over the return horizon (1, 3 and 5 days), label threshold (0.5 to 2.0 standard deviations), decay half-life and news lag.
\item Use time-stamped news to treat after-close items exactly, and extend the trading test with position sizing, borrow costs and slippage.
\end{itemize}

\section*{Reproducibility}
All data files, code, results and the scripts that build every table and figure are available at \url{https://github.com/sohomiitb/News_Impact_Forecasting}. Running \texttt{python src/run\_experiments.py} followed by \texttt{python src/make\_report.py} regenerates the results in this paper.

\section*{Funding Statement}
This research received no external financial support.

\bibliographystyle{plainnat}

\end{document}